\documentclass[aps,10pt,prd,preprintnumbers,amsmath,amssymb,nofootinbib,superscriptaddress,a4paper,showpacs,twocolumn]{revtex4-1}

\usepackage[breaklinks]{hyperref}
\usepackage[utf8]{inputenc}
\usepackage[OT2,T1]{fontenc}
\usepackage{graphicx}
\usepackage{color}
\usepackage{amsfonts,amsthm}
\usepackage{bm,bbm}
\usepackage{amsmath,amssymb,amsfonts} 

\newcommand{\be}{\begin{equation}}
\newcommand{\ee}{\end{equation}}
\newcommand{\ba}{\begin{eqnarray}}
\newcommand{\ea}{\end{eqnarray}}
\def\bs{\begin{subequations}}
\def\es{\end{subequations}}
\def\a{\alpha}

\def\g{\gamma}
\def\la{\lambda}

\def\om{\omega}

\def\s{\sigma}

\def\cP{\mathcal{P}}

\def\cV{\mathcal{V}}

\def\ds{d_{\rm S}}
\def\dh{d_{\rm H}}

\def\dch{d_{\mathbbm{C}{\rm H}}}
\def\dcs{d_{\mathbbm{C}{\rm S}}}

\newcommand{\Eq}[1]{(\ref{#1})}
\def\com{\color{magenta}}
\def\cob{\color{blue}}
\newcommand{\book}[5]{\emph{#1} (#2, #3, #4, #5)}
\newcommand{\books}[4]{\emph{#1} (#2, #3, #4)}
\newcommand{\oarX}[1]{\href{http://arxiv.org/abs/#1}{{\ttfamily\com arXiv:#1}}}
\newcommand{\arX}[1]{\href{http://arxiv.org/abs/#1}{{\ttfamily\com arXiv:#1}}}
\newcommand{\doin}[6]{\href{http://dx.doi.org/#1}{{\cob #2 #3 {\bf #4}, #5 (#6)}}}
\newcommand{\doinn}[5]{\href{http://dx.doi.org/#1}{{\cob #2 {\bf #3}, #4 (#5)}}}
\newcommand{\doij}[5]{\href{http://dx.doi.org/#1}{{\cob #2 #3 (#5) #4}}}

\newcommand{\procm}[6]{in \emph{#1}, ed.\ by #2 (#3, #4, #5, #6)}
\newcommand{\procsinm}[5]{in \emph{#1}, ed.\ by #2 (#3, #4, #5)}
\newcommand{\tia}[1]{{#1},}

\def\lp{\ell_{\rm Pl}}

\def\rme{\text{e}}
\def\rmd{\text{d}}
\def\rmi{\text{i}}

\begin{document}

\title{Complex dimensions and their observability}

\author{Gianluca Calcagni}
\email{calcagni@iem.cfmac.csic.es}
\affiliation{Instituto de Estructura de la Materia, CSIC, Serrano 121, 28006 Madrid, Spain}

\date{May 3, 2017}

\begin{abstract}
We show that the dimension of spacetime becomes complex-valued when its short-scale geometry is invariant under a discrete scaling symmetry. This characteristic can generically arise in quantum gravities, for instance, in those based on combinatorial or multifractal structures or as the partial breaking of continuous dilation symmetry in any conformal-invariant theory. With its infinite scale hierarchy, discrete scale invariance overlaps with the traditional separation between ultraviolet and infrared physics and it can leave an all-range observable imprint, such as a pattern of log oscillations and sharp features in the cosmic microwave background primordial power spectrum.
\end{abstract}


\preprint{\doin{10.1103/PhysRevD.96.046001}{Phys.\ Rev.}{D}{96}{046001}{2017} \hspace{10.5cm} \arX{1705.01619}}

\maketitle

\section{Introduction}

Detecting a signature of quantum gravity is perhaps one of the most exciting things that could happen in the physics of the 21st century. The quest for a theory beyond general relativity and the Standard Model of particles, unifying gravitation and quantum mechanics, has been going on for decades and it is not planning to stop anytime soon. This search does not lack diversity: there are about a dozen major scenarios trying to quantize gravity independently \cite{Ori09,Fousp,BaMcA,CQC} and, in some cases, also to unify all fundamental forces of Nature. Names such as string theory, group field theory, asymptotic safety, loop quantum gravity, causal dynamical triangulations, noncommutative spacetimes, and several others are familiar to the expert, but they may evoke only vague images to those not working in the field. Part of the reason is that we have no empirical evidence in favor of any of these proposals, either because the theory has not been developed to the point of producing robust observable predictions, or effects are negligible and undetectable, or, finally, simply because we have not reached enough experimental sensitivity. 

Having more than one candidate theory creates the added problem of fragmentation of efforts, but there is regularity in this heterogeneity. In all known cases, the dimension of spacetime changes with the probed scale (dimensional flow) \cite{tH93,Car09,fra1,revmu}. The energy $E_*$ at which the dimension differs considerably from the observed value $D=4$ is typically too large (of order of the grand-unification or Planck scale, $E_*\gtrsim 10^{15}\!-\!10^{19}\,\text{GeV}$) to be probed directly, which means that running dimensions may remain an academic curiosity, a beautiful universal feature of quantum gravity with no applications. But is that so? In this paper, we argue for a negative answer with important phenomenological consequences. We show that, in theories where the short-scale spacetime geometry has discrete symmetries in the ultraviolet (UV), discreteness can originate a \emph{long-range} modulation of the spacetime geometry in the form of logarithmic oscillations. These ripples extend at observable scales much larger than the Planck length $\lp$. This phenomenon is associated with the arising of a complex-valued dimension, a concept that, at first, might sound very unphysical even to the most liberal theoretician. However, we will see that complex dimensions, an old acquaintance of fractal geometry \cite{BGM1,DDSI,BGM2,DILu,BFSTV,ErEc,NLM,LvF}, are nothing but the frequencies $n\om$ of the long-rage modulation generated by UV discreteness, which could be observed directly in the sky as a pattern of sharp spikes in the cosmic microwave background (CMB) spectrum. We will discuss this general quantum-gravity prediction in the special case of multifractional theories, improving on very preliminary (single harmonic, no infinite hierarchy) results \cite{revmu,frc14}.

\section{Discrete scale invariance (DSI) and log oscillations} 

To begin with, we describe how a long-range modulation arises from a discrete symmetry and how it relates to dimensional flow in quantum gravity. This section is more technical than the rest, but not overly so and the payback in terms of physical insight is worth the effort. Let us derive the general form of this modulation from a scaling argument well known in critical systems \cite{NiLe} and fractal geometry \cite{BGM2}, but not employed so far in the literature of quantum gravity (but see \cite{Akk12}). Consider the half line $z\geq 0$. Scale invariance, also known as self-similarity, is the symmetry of a function $f(z)$ under a coordinate arbitrary rescaling, $f(\la z)= c f(z)$ for any $\la$, where $c$ is a constant. The solution of this equation is a power law $f(z)\propto z^\a$, so that $c=\la^\a$. Critical systems usually have scale-invariant properties governed by power laws, near the critical point of a continuous phase transition; then, $\a$ is called a critical exponent. On the other hand, $f$ is invariant under a \emph{discrete scaling symmetry} if
\be\label{f1}
f(\la_\om z)=\la_\om^\a [f(z)-g(z)]
\ee
for some \emph{fixed} $\la_\om$ and $\a$, where $g(z)$ is a regular function analytic at all points $z\geq 0$. The ordinary continuous scaling symmetry corresponds to an arbitrary dilation factor $\la$. Focusing our attention on the nonanalytic behavior $f-g$ close to the ``critical'' point $z=0$, we can ignore $g$ as a first approximation. In quantum gravity, this would correspond to consider the UV behavior of $f$ at scales $z=\ell/\ell_*\ll 1$, where $\ell_*$ is a reference scale separating the UV from the infrared (IR) and $\ell$ is the probed length. The solution of \Eq{f1} with $g(z)=0$ is
\be\label{fx}
f(z)=z^\a F_\om(z)\,,\qquad F_\om(\la_\om z)=F_\om(z)\,.
\ee
Rewriting $F_\om(z)=G(\ln z/\ln\la_\om)$, the scale invariance of $F_\om(z)$ translates into a logarithmic periodicity of $G$ with period 1 \cite{BGM1,DDSI,BGM2,DILu,BFSTV,ErEc}. 
 
Restoring the regular part $g\neq 0$, the general solution of \Eq{f1} is the Weierstrass-type function
\be\label{fz}
f(z)=g(z)+\sum_{n=1}^{+\infty} \la_\om^{-\a n}g(\la_\om^n z)\,.
\ee
For a certain range of $\la_\om$ and $\a$, $f$ is nowhere differentiable; $g=\cos$ is the original case considered by Weierstrass in 1872. Equation \Eq{fz} is related to Eq.\ \Eq{fx} by the so-called zeta function, which we will need also to determine the dimension of spacetime. Let $z\geq 0$. The zeta function of a function $f(z)$ is defined by the Mellin transform $\zeta_f(s):=\int_0^{+\infty}\rmd z\,z^{s-1}f(z)/\Gamma(s)$. Applying this to a function with a DSI, from Eq.\ \Eq{f1} and inverting with respect to $\zeta_f$ one has
\be\label{zetaf}
\zeta_f(s)=\frac{\la_\om^{s+\a}\zeta_g(s)}{\la_\om^{s+\a}-1}\,.
\ee
Since $g(z)$ is regular for $z>0$, the poles of $\zeta_g$ take only negative integer values $s_j=-j$, $j\in\mathbbm{N}^+$, and contribute only to the regular part of $f(z)$ \cite{Sor98,GlSo}. The other poles $s_l$ of $\zeta_f$ are the solutions of the complex-valued equation $\la_\om^{s_l+\a}=\rme^{-2\pi\rmi l}$, where $l\in\mathbbm{Z}$. Redefining $\la_\om=:\exp(-2\pi/\om)$, then 
 $s_l=-\a+\rmi l \om$. Consistently, plugging Eq.\ \Eq{zetaf} into the inverse Mellin transform $f(z)=(2\pi\rmi)^{-1}\int_{\g-\rmi\infty}^{\g+\rmi\infty}\rmd s\,z^{-s}\Gamma(s)\,\zeta_f(s)$ and using Cauchy residue theorem, one gets
\be
f(z)=z^\a\sum_{l=-\infty}^{+\infty}\Gamma(s_l)\zeta_g(s_l)\rme^{-\rmi l \om\ln z}=:z^\a F_\om(z),\label{fx2}
\ee
exactly reproducing the profile \Eq{fx}. Furthermore, if $f$ is real (as in fractals), then one rearranges the sum to get
\be\label{fomz}
F_\om(z)=A_0+\sum_{n=1}^{+\infty}[A_n\cos(n\om\ln z)+B_n\sin(n\om\ln z)]\,,
\ee
where $A_0=\Gamma(-\a)\zeta_g(-\a)$ and $(A_n\mp\rmi B_n)/2=\Gamma(s_{\pm n})\zeta_g(s_{\pm n})$.

Discrete scale invariance is an ubiquitous feature of many physical situations involving disorder, growth and rupture \cite{SJAMS,JoS}, turbulence \cite{Nov66,Nov90,JSH}, financial crashes \cite{SJB,JoS2,LRSo}, earthquake precursory seismicity \cite{SoSa1,SSS,HuSS}, as well as chaotic, spin, critical, and general hierarchical systems \cite{Erz97,Sor98}. DSI is especially important in fractal geometry \cite{BGM1,DDSI,BGM2,NLM}. Deterministic (i.e., exactly self-similar) fractals are described by fractional measures with log oscillations, while random fractals (randomized scaling ratio $\la$ at each iteration of the set) are obtained by averaging over oscillations \cite{NLM}. This is the fundamental basis from which multifractional quantum theories of matter fields and gravity were formulated \cite{revmu}.

\section{Spacetime dimensions} 

Quite surprisingly, the log-periodic modulation factor $F_\om$ in Eq.\ \Eq{fx2} arises in all quantum gravities where the Hausdorff and/or spectral dimension vary with the scale \cite{first}. Throughout this work, we do not distinguish between space and spacetime, since the time direction is usually Euclideanized when considering dimensions.

The \emph{Hausdorff dimension} $\dh(\ell):=\rmd\ln\cV/\rmd\ln\ell$ is the scaling of the volume $\cV(\ell)$ of a $D$-ball (or a $D$-cube) with radius (respectively, edge size) $\ell$. If $\dh={\rm const}$, then $\cV(\ell)\propto \ell^{\dh}$ and, in ordinary space, $\dh=D$ coincides with the topological dimension. If the Hausdorff dimension $\dh(\ell)$ is scale-dependent, then the volume scales as \cite{first}
\be\label{cV}
\cV(\ell) =\ell^D\left[1+\sum_{n=1}^{+\infty}\frac{1}{\a_n}\left(\frac{\ell}{\ell_n}\right)^{D(\a_n-1)}F_n(\ell)\right],
\ee
where $F_n(\ell)= b_{0,n}+b_{n}({\ell}/{\ell_\infty})^{-\rmi n\om}+b_{-n}({\ell}/{\ell_\infty})^{\rmi n\om}$, $b_{0,n}$ and $b_{\pm n}$ are complex, we introduced a scale hierarchy $\{\ell_\infty,\ell_{n+1}<\ell_n\}$ for dimensional reasons, and $\a_{n+1}>\a_n$ and $\om$ are real parameters. For a monotonic dimensional flow with UV Hausdorff dimension smaller than in the IR, $b_{0,1}>0$ and $0\leq\a_n<1$. The details of the model determine all these parameters, but the form \Eq{cV} of $\cV(\ell)$ is universal. In particular, in the UV ($\ell\ll\ell_*\equiv\ell_1$) $\dh^{\rm uv}\simeq D\a$ (where $\a\equiv\a_1$), while $\dh^{\rm IR}\simeq D$. Equations \Eq{fx2} and \Eq{cV} agree under the identification $f=\cV/{\ell^D}-1$, $z={\ell}/{\ell_\infty}$, $\a_n=\a$, and $\ell_n=\ell_*$. [One should not confuse the regular part $g(z)$ with the term $\ell^D$ in Eq.\ \Eq{cV}: in quantum gravity, DSI is an approximate symmetry valid only in the UV.] In this configuration, there are only two scales $\ell_\infty$ and $\ell_*$; the general multiscale case (many $\a_n$ and $\ell_n$) corresponds to a multi-DSI geometry.

The \emph{spectral dimension} $\ds(\ell):=-\rmd\ln\cP/\rmd\ln\ell$ is the scaling of the inverse of the return probability $\cP(\ell)$ with the probed scale $\ell$, a quantity related to the probability that a test particle propagates back to a given initial point in spacetime (see \cite{CMNa} for a review). Thanks to the formal duality $\cV\leftrightarrow1/\cP$, one can translate all results for the Hausdorff dimension to the spectral dimension \cite{first}. If $\ds={\rm const}$, then $\cP(\ell)\propto \ell^{-\ds}$, while the general form of $\ds(\ell)$ is obtained from the definition and the inverse of Eq.\ \Eq{cV}. Equation \Eq{fx2} is reproduced for $f=(\cP\ell^D)^{-1}-1$.

Unless one considers highly nontrivial states of quantum geometry, the volume $\cV$ and the return probability $\cP$ (which, in quantum gravity, can be replaced by the expectation values $\langle\hat\cV\rangle$ and $\langle\hat\cP\rangle$ of, respectively, the volume and return-probability operator \cite{COT2,COT3}) are real-valued and so are $\dh$ and $\ds$. Then, $b_{n}=b_{-n}^*$ and the modulation factor in Eq.\ \Eq{cV} becomes a superposition $F_\om(\ell)=\sum_n F_n(\ell)$ of logarithmic oscillations:
\be\label{logs}
F_n(\ell)=b_{0,n}+A_n\cos\left(n\om\ln\frac{\ell}{\ell_\infty}\right)+B_n\sin\left(n\om\ln\frac{\ell}{\ell_\infty}\right),
\ee
in agreement with Eq.\ \Eq{fomz} with $A_0=\sum_nb_{0,n}$. Self-similar fractals are characterized by a real-valued measure and, hence, by log oscillations \cite{LvF}.

As anticipated, the log-oscillating modulation $F_\om(\ell)$ is \emph{long-range} because it is associated with an infinite scale hierarchy $\{\ldots,\la_\om^{-2}\ell_\infty,\la_\om^{-1}\ell_\infty,\ell_\infty,\la_\om\ell_\infty,\la_\om^2\ell_\infty,\ldots\}$ in geometric progression. While the traditional separation between UV and IR, typical of quantum field theory, is realized by the power-law part in Eq.\ \Eq{cV}, the DSInvariant part $F_\om$ is insensitive to this dichotomy and modulates infinitely many intermediate scales $\la_\om^{\pm n}\ell_\infty$ spanning all ranges. This constitutes a major departure from the standard quantum-gravity wisdom about a UV/IR divide, inasmuch as the impact of the above scale hierarchy can be much heavier and more characteristic than the imprint usually looked for in physical observables. Quantum gravity may be more accessible than believed.

Only some quantum gravities happen to have a variable $\dh$, but almost all have a variable $\ds$; see Refs.\ \cite{CQC,revmu} for a case-by-case summary. A scale-dependent dimension implies the existence of at least one characteristic scale $\ell_*$, or in some cases a whole scale hierarchy $\{\ell_n\}$. For this reason, quantum gravities are said to describe \emph{multiscale} geometries. On the other hand, only multifractional spacetimes have explicit log oscillations (both in the volume and the return probability) \cite{revmu,frc2}, while no hint of them has been so far recognized in other proposals. The present work aims to revise this claim.

The above definition of $\ds$ has been recently replaced, in quantum gravity, by a less intuitive but much more appealing technique, adapted from the $\ds={\rm const}$ case of fractal geometry \cite{BGM1,DDSI,BGM2,LvF}, that identifies the values of the spectral dimension in the plateaux $\ds\simeq{\rm const}$ of the profile $\ds(\ell)$ with the poles of the spectral zeta function \cite{ArCa3}. Here we apply, for the first time, the same procedure to $\dh$ and show that the Hausdorff dimension of spacetime also has an interpretation in terms of the poles of the zeta function associated with the volume. This will allow us to discover, so to speak, the existence of complex dimensions. The reader can immediately recast all the results for the case of $\ds$, by replacing the volume $\cV$ with the inverse return probability $1/\cP$.

Recall that Eq.\ \Eq{fx2} is a log-periodic series with coefficients given by the poles $s_l=-\a+\rmi l \om$ of the zeta function. Identifying $f$ with $\cV/\ell^D-1$ and $z\propto\ell$, in $D=1$ topological dimensions this expression determines the UV Hausdorff dimension $\dh=\a$ and the \emph{complex Hausdorff dimension}
\be\label{dch}
\dch=\om\,.
\ee
This result can be guessed also by noting that log oscillations come from the combination of complex powers $x^{\a\pm\rmi\om}$, immediately leading to the identification of the exponent as a sort of complex-valued dimension. Higher-order harmonics $|l|>1$ give rise to higher dimensions $|l|\dch$.

For the non-DSInvariant function \Eq{cV}, there will be two families of poles, $s_l=-D+\rmi l \om$ and $s_m=-D\a+\rmi m \om$, giving the IR and UV Hausdorff dimensions $\dh\simeq D,D\a$ and the same complex dimension \Eq{dch}. [The IR pole can be obtained after regularizing the Mellin transform or, exactly, in theories where the dimensional flow $\dh(\ell)$ is known analytically at all scales, such as in multifractional spacetimes; see \cite{ArCa3} for the case of $\ds$.] Thus, $\dch$ is independent of the topological dimension $D$. Similarly, the complex spectral dimension $\dcs$ will coincide with the frequency $\om_{\rm S}$ in the return probability (which may differ with respect to the frequency $\om$ in the volume).

Thus, the Hausdorff dimension of the plateaux of the dimensional flow of a Euclideanized spacetime is minus the real part of the poles of the zeta function of the spacetime measure (volumes), and the complex Hausdorff dimension is the smallest nonzero imaginary part:
\be
s_{i,l}=-\dh^{(i)}+\rmi l\,\dch\,,\qquad \text{$i$th plateau}\,.
\ee
The poles of the zeta function of the return probability have the same structure up to a sign, $s_{i,l}=\ds^{(i)}+\rmi l\dcs$, possibly with an extra factor of $1/2$ in the real part if another scale variable $\s\propto\ell^2$ is used instead \cite{ArCa3}. This concludes the proof that, quite generically, a spacetime with dimensional flow and a UV DSI will have a \emph{variable} Hausdorff and/or spectral dimension $\dh(\ell),\ds(\ell)$ and a \emph{constant} complex Hausdorff and/or spectral dimension $\dch,\dcs$, corresponding to the frequency of the log oscillations in the volume and/or return probability.

\section{DSI in quantum gravity} 

Complex dimensions, also called complex critical exponents, appear whenever there is a DSI. The simplest way to realize DSI in quantum gravity is when spacetime (pre)geometry has a fundamentally discrete structure similar to that of hierarchical lattices \cite{Erz97}. This structure, explicit in multifractional theories \cite{revmu}, is the most general one realizing dimensional flow in quantum gravity \cite{first}. Then, logarithmic oscillations arise as an expression of a DSI in the deep UV. However, the combinatorial structure of scenarios such as loop quantum gravity, spin foams, and group field theory can be much less regular than a hierarchical lattice and complex dimensions may arise but only in special situations. This possibility is still under investigation, although preliminary results indicate a complex-valued spectral dimension in certain superpositions of highly quantum states of geometry \cite{COT2}. 

A more flexible scenario we propose here is that a DSI appears spontaneously from the soft breaking of continuous scale invariance $x^\mu\to\la x^\mu$, a subsymmetry of conformal-invariant field theories. For example, if the scaling ratio $\la$ were the expectation value of a field and if such field took discrete values $\la\to\la_\om^n$ in geometric progression, we would get a DSI without any underlying hierarchical structure. To achieve this, and inspired by flux compactification in low-energy string theory \cite{CQC,GKP}, we could take $p$-forms $A_p$ on a compact subspace $\Gamma_{p+1}$, which produce quantized fluxes $\int_{\Gamma_{p+1}}\rmd A_p= l q$, where $q=:2\pi/\om$ is the $p$-form magnetic charge. An effective action $S[g_{\mu\nu},A_p]$ characterized by a conformally coupled metric $\rme^{-2\int\rmd A_p}g_{\mu\nu}$ is associated with a coordinate dilation with arbitrary scaling ratio $\la\sim \rme^{-\int\rmd A_p}$. However, upon quantizing the flux, one forces the system to a discretized conformal coupling $\la_\om^{2l}g_{\mu\nu}$. Then, a DSI and the inevitable log oscillations are generated dynamically. Also, different forms yield different fluxes and frequencies $\om$, whose superposition can give rise to a log-quasi-periodic structure \cite{Sor98,SJAMS} with a richer behavior. A more precise treatment should be able to make this mechanism rigorous, at least in a Euclidean setting. Theories in Lorentzian signature are more delicate to deal with, since complex exponents lead to exponentially divergent correlation functions after Wick rotation \cite{SaSo}. This might suggest that a UV-finite theory of quantum gravity admitting a well-defined analytic continuation to and from imaginary time should have no log oscillations in the time direction.

\section{Can we observe complex dimensions?} 

In general, complicating a simple rigid DSInvariant structure, for instance by disorder effects, does not spoil log oscillations, and complex exponents are found to be robust against small perturbations \cite{Sor98}. This can be relevant in quantum gravity, where coarse graining by low-resolution instruments \cite{frc2} (i.e., averaging over a log-period \cite{NLM}) or highly nontrivial superpositions of geometry states could randomize a UV DSI and wipe it out. However, in a multiscale geometry this phenomenon of destructive interference does not cancel the log oscillations completely and we may wonder whether their footprint is observable. Here we can gain an insight by the following remark.

In all known examples in critical, complex, and fractal systems, the amplitudes $A_n$ and $B_n$ in Eq.\ \Eq{logs} decay with $n$ either exponentially ($A_n,B_n\sim\rme^{-\g n}$, as in the Potts model with antiferromagnetic interactions or in walks on DSInvariant fractals) or as a power law ($A_n,B_n\sim n^{-u}$, as in the Potts model with ferromagnetic interactions) \cite{Sor98,GlSo}. As a consequence, the first harmonics $n=1,2,\ldots, n_{\rm max}$ are enough to fit experimental data. This loosely justifies the use of the first harmonic in multifractional theories \cite{revmu,frc14}, but another possibility is to consider several harmonics simultaneously and constrain the microstructure of quantum geometry from their amplitudes. In fact, the above two classes of amplitude behavior (exponential or power law) correspond to different interaction properties among the sites of a discrete structure, as illustrated by Potts models of interacting spins on lattices \cite{GlSo}. If higher-order log oscillations could be constrained efficiently, then one could determine their suppression laws for $A_n$ and $B_n$ and gain a valuable insight about the details of the spacetime structure (fundamental or effective) originating it, for instance how spin labels distribute in the pregeometric combinatorial complexes of some discrete quantum gravities. One can investigate whether present-generation CMB data (such as those of the last \textsc{Planck} release) are sensitive enough to this higher-order effect. We can get a flavor of the outcome by considering the parametrization
\be\label{Aa}
A_n=a_n \frac{\rme^{-\g n}}{n^u},\qquad B_n=b_n \frac{\rme^{-\g n}}{n^u},
\ee
which is the real-valued analogue of the complex amplitudes classified in Ref.\ \cite{GlSo}. The coefficients $a_n,b_n$ and the parameters $\g$ and $u$ are determined by the Mellin transform of $g$, as noticed below Eq.\ \Eq{fomz}. 

\begin{figure}[ht]
\centering
\includegraphics[width=8.6cm]{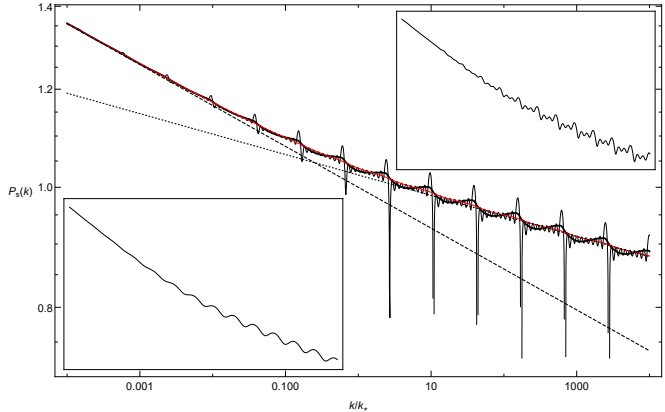}
\caption{\label{fig1} Main plot: log-log plot of the multifractional primordial scalar power spectrum for $n_{\rm max}=9$ and $\g=0=u$ (thin black curve), $\g=0$ and $u=2$ (thick black curve), and $\g=2$ and $u=0$ (red curve), compared with the standard spectrum (dashed line) and the deep-UV trend $k^{\a(n_{\rm s}-1)}$ (dotted line). Insets: case $\g=0=u$ with $n_{\rm max}=1$ (lower plot) and $n_{\rm max}=3$ (upper plot). The other parameters are $a_n=b_n=0.2$, $k_\infty=1/\lp\approx 10^{57}\text{Mpc}^{-1}$ \cite{revmu,frc14}, $k_*=1$, $\om=2\pi\a/\ln 2$ \cite{frc14}, $n_{\rm s}=0.967$ \cite{P1520}, $\a=1/2$.}
\end{figure}

We apply Eq.\ \Eq{Aa} to a cosmological example of quantum-gravity-related phenomenology, the inflationary scalar power spectrum of the multifractional theory with $q$-derivatives. In this fundamentally multiscale spacetime \cite{revmu,frc14}, dispersion relations are deformed, which results in a modification of the primordial scalar perturbation spectrum $P_{\rm s}$. For an isotropic binomial measure with a UV intrinsic DSI, the massless dispersion relation $k^2=0$ is deformed into $p^2(k)=p_\mu(k^\mu) p^\mu(k^\mu)=0$ (sum only over $p$ indices), where for each direction (no index summation)
\ba
p^\mu(k^\mu)&=&\frac{1}{\cV_{D=1}[(k^\mu)^{-1}]}\nonumber\\
&=&k^\mu\left[1+\frac{1}{\a}\left(\frac{k^\mu}{k_*}\right)^{1-\a}\sum_n F_n(k^\mu)\right]^{-1},
\ea
where $F_n(k)$ is the momentum-space analogue of Eq.\ \Eq{logs} and all the scales $\ell_n^{-1}$ and $\ell_\infty^{-1}$ are replaced by momentum (energy) scales $k_n,k_\infty$ (here there is only one $k_1\equiv k_*$). In particular, the propagation of all classical and quantum particles is affected by DSI, for instance as a log-periodic modulation of the propagation speed. In the context of inflation, having composite momenta $p^\mu(k^\mu)$ produces a bending of the spectrum at small scales superposed to a modulation pattern. While the standard scalar spectrum is parametrized, to leading order in a slow-roll approximation, as $P_{\rm s}(k)\propto k^{n_{\rm s}-1}$, where $k=|{\bf k}|$ is the spatial comoving wavenumber and $n_{\rm s}$ is the scalar spectral index, in the theory with $q$-derivatives this parametrization would be inadequate because the slow-roll approximation can be relaxed and the spectral index be far away from 1 in certain regimes. Thus, it is more convenient to adopt, up to numerical factors, the form
\be
P_{\rm s}(k)\propto[p(k)]^{n_{\rm s}-1}, 
\ee
where $p(k)=|{\bf p}|$. This $P_{\rm s}(k)$ is the exact extension to all harmonics of the first-harmonic approximate spectrum found in Ref.\ \cite{frc14}. Figure \ref{fig1} shows several types of behaviour up to $n_{\rm max}$ harmonics. If $A_n,B_n$ are constant, then when $n_{\rm max}$ increases oscillations become \emph{spikes} at logarithmic intervals. When $\g$ increases, amplitudes decrease and they virtually disappear into tiny ripples already for $\g\simeq 1$. When $u$ increases, the plot becomes festooned already for $u\simeq 1$. 

The sensitivity of CMB data \cite{P1520} to features \cite{CHP}, especially sharp ones such as the above spikes, could mark the imprint we are looking for. The temperature spectrum \cite{KHSS}, temperature non-Gaussianity \cite{GLHM,AGHSS,GPS}, and polarization \cite{MHD} can all be scanned for this purpose. Usually, fine features are inserted \emph{ad hoc} in the power spectrum \cite{CHP,INY} or generated by inflationary dynamics \cite{CHP,KYS,Pal14,HSSS}, while in the present case they arise purely from geometry. Since features can even improve the fit of \textsc{Planck} 2015 data \cite{HSSS}, the issue at stake here is not just how to constrain the parameters of one or more quantum-gravity models, but also whether these models can account for some details of the CMB multipole spectrum better than the standard $\Lambda$CDM model of general relativity. Following a likelihood analysis similar to that of Ref.\ \cite{frc14}, we will report on this in a future publication.

\section*{Acknowledgments} 

The author is under a Ram\'on y Cajal contract, is supported by the I+D grant FIS2014-54800-C2-2-P, and thanks S.\ Kuroyanagi and L.\ Modesto for discussions.

\bigskip
\noindent\emph{Note added.---}After submission, \cite{BOA} appeared where the authors discussed a phase transition between a continuous and a discrete scale invariance in Lifshitz-type Hamiltonian models. After some elaboration, these results can have a direct application to Ho\v{r}ava--Lifshitz gravity and corroborate the main claim of this paper, providing another and important explicit example of how discrete scale invariance (and complex dimensions) can emerge naturally in quantum gravities. This example adds to the two cases discussed here, multifractional theories and a generic conformal-invariant theory with $p$-forms coupled to the metric.

\end{document}